\pdfoutput=1 
\documentclass[conference]{IEEEtran}
\usepackage{makecell}
\usepackage{hyperref}
\usepackage{array}
\usepackage{graphicx,amssymb,amsmath}
\usepackage{multicol}
\usepackage[noadjust]{cite}
\usepackage{setspace}
\usepackage{graphicx}
\usepackage{float}
\usepackage {url}
\usepackage{stfloats}
\usepackage{amsthm,pifont}
\usepackage{flushend}
\usepackage{cases,subeqnarray}
\usepackage{bm,multirow,bigstrut}
\usepackage{amsmath, amsthm, amssymb}
\usepackage{textcomp}
\usepackage{latexsym,bm}
\usepackage{booktabs}
\usepackage{xcolor}
\usepackage{mathtools}
\usepackage{dsfont}
\usepackage{extarrows}
\usepackage{epsfig}
\usepackage{subcaption}
\usepackage{epstopdf}
\usepackage[noend]{algpseudocode}
\usepackage{makecell}
\usepackage{colortbl}
\usepackage{booktabs}
\usepackage{graphicx} 
\usepackage{array}    
\usepackage{hyperref} 
\usepackage{booktabs}
\usepackage{enumitem}
\usepackage{diagbox}
\usepackage{tikz}
\usetikzlibrary{trees,shadows.blur}
\usepackage{forest}
\usepackage{algorithm}
\usepackage{algorithmicx}
\usepackage{threeparttable}
\theoremstyle{plain}

\theoremstyle{plain}

\usepackage{amsmath}

\newcommand{\ignore}[1]{{{\color{yellow} }}}
\definecolor{blue-green}{rgb}{0.0, 0.87, 0.87}
\IEEEoverridecommandlockouts

\begin{document}
\setlength{\columnsep}{0.21in}

\title{Maximum-Likelihood Estimation Based on Diffusion Model For Wireless Communications
}

\author{Changyuan Zhao, Jiacheng Wang, Ruichen Zhang, Dusit Niyato, \textit{Fellow, IEEE},  \\
Dong In Kim, \textit{Fellow, IEEE}, Hongyang Du*
\thanks{C. Zhao is with the College of Computing and Data Science, Nanyang Technological University, Singapore, and CNRS@CREATE, 1 Create Way, 08-01 Create Tower, Singapore 138602 (e-mail: zhao0441@e.ntu.edu.sg).
}
\thanks{J. Wang, R. Zhang, and D. Niyato are with the College of Computing and Data Science, Nanyang Technological University, Singapore (e-mail: jiacheng.wang@ntu.edu.sg; ruichen.zhang@ntu.edu.sg; dniyato@ntu.edu.sg).}
\thanks{D. I. Kim is with the Department of Electrical and Computer Engineering, Sungkyunkwan University, Suwon 16419, South Korea (email:dongin@skku.edu).}
\thanks{H. Du is with the Department of Electrical and Electronic Engineering, University of Hong Kong, Pok Fu Lam, Hong Kong (e-mail: duhy@eee.hku.hk).}
}


\author{
 \thanks{This research is supported by the Singapore Ministry of Education (MOE) Tier 1 (RG87/22 and RG24/24), the NTU Centre for Computational Technologies in Finance (NTU-CCTF), and the RIE2025 Industry Alignment Fund - Industry Collaboration Projects (IAF-ICP) (Award I2301E0026), administered by A*STAR; in part by the National Research Foundation of Korea (NRF) Grant funded by the Korean Government (MSIT) under Grant 2021R1A2C2007638.}
    \IEEEauthorblockN{
    Changyuan Zhao\IEEEauthorrefmark{2}\IEEEauthorrefmark{3}, 
    Jiacheng Wang\IEEEauthorrefmark{2}, 
    Ruichen Zhang\IEEEauthorrefmark{2}, 
    Dusit Niyato\IEEEauthorrefmark{2}, 
    Dong In Kim\IEEEauthorrefmark{1}, 
    Hongyang Du\IEEEauthorrefmark{4}\thanks{*Corresponding author: Dong In Kim (email: dongin@skku.edu)}
    }   
    \IEEEauthorblockA{\IEEEauthorrefmark{2}College of Computing and Data Science, Nanyang Technological University, Singapore 639798, Singapore}
    \IEEEauthorblockA{\IEEEauthorrefmark{3}CNRS@CREATE, 1 Create Way, 08-01 Create Tower, Singapore 138602, Singapore}
    \IEEEauthorblockA{\IEEEauthorrefmark{1}Department of Electrical and Computer Engineering, Sungkyunkwan University, Suwon 16419, South Korea}
    \IEEEauthorblockA{\IEEEauthorrefmark{4}Department of Electrical and Electronic Engineering, University of Hong Kong, Pok Fu Lam, Hong Kong}
}

\maketitle
\vspace{-1cm}

\begin{abstract}

Generative Artificial Intelligence (GenAI) models, with their powerful feature learning capabilities, have been applied in many fields. In mobile wireless communications, GenAI can dynamically optimize the network to enhance the user experience. Especially in signal detection and channel estimation tasks, due to digital signals following a certain random distribution, GenAI models can fully utilize their distribution learning characteristics. For example, diffusion models (DMs) and normalized flow models have been applied to related tasks. However, since the DM cannot guarantee that the generated results are the maximum-likelihood estimation points of the distribution during the data generation process, the successful task completion rate is reduced. Based on this, this paper proposes a Maximum-Likelihood Estimation Inference (MLEI) framework. The framework uses the loss function in the forward diffusion process of the DM to infer the maximum-likelihood estimation points in the discrete space. Then, we present a signal detection task in near-field communication scenarios with unknown noise characteristics. In experiments, numerical results demonstrate that the proposed framework has better performance than state-of-the-art signal estimators.


\end{abstract}
\IEEEpeerreviewmaketitle

\section{Introduction}\label{intro}

The Generative Artificial Intelligence (GenAI) model is an unsupervised learning model that aims to learn the distribution characteristics of samples in order to generate similar high-quality content. Due to its remarkable generation ability, the GenAI methods, particularly diffusion models (DMs) and large language models, have found widespread applications in areas such as text generation, image generation, and audio generation \cite{zhao2024enhancing, qu2024performance}.  
Therefore, researchers have increasingly recognized its potential and have begun applying it in a broader range of disciplines.
In wireless communication areas, the transition from 5G to 6G wireless networks signifies a shift to a more sophisticated, immersive, and integrated network ecosystem \cite{wang2024tutorial}. To address the demanding requirements of 6G, the deployment of GenAI leverages the potent capabilities of GenAI to optimize vehicular network configurations dynamically, enhance signal processing, and improve intelligent context-aware services \cite{du2024enhancing, zhao2024generative, wang2024generative}. 

Within mobile wireless communication tasks, signal detection and channel estimation are classic and crucial processes aiming to ensure the integrity of communication and improve communication quality \cite{huang2018deep}.
Taking signal detection as an example, researchers adopted the maximum a posteriori (MAP) estimation to compute the detected signal solution, which can be written as follows:
\begin{equation}
\begin{split}
    \widehat{\bm{x}}_{MAP} = \arg\max_{\bm{x}\in\mathcal{X}}p(\bm{x}\vert\bm{y})
     = \arg\max_{\bm{x}\in\mathcal{X}}p(\bm{y}\vert\bm{x})p(\bm{x}),
\end{split}
\end{equation}
where $\mathcal{X}$ represents the set of all possible signal vectors, and $p(\cdot)$ denotes probability density function.
When there is no prior knowledge of the transmitted signal $\bm{x}$, the MAP estimation is equivalent to the maximum-likelihood estimation (MLE), which can be formulated as follows \cite{he2021learning}:
\begin{equation}
\label{eq:mle}
\begin{split}
    \widehat{\bm{x}}_{MAP} =  \widehat{\bm{x}}_{MLE} &= \arg\max_{\bm{x}\in\mathcal{X}}p(\bm{y}\vert\bm{x})\\
    & = \arg\max_{\bm{x}\in\mathcal{X}}p_{n}(\bm{y}-\bm{Ax}),
\end{split}
\end{equation}
where $\bm{A}$ denotes the channel matrix, and $p_{n}(\cdot)$ is an independent and identically distributed (i.i.d) measurement of noise.

Generally, when assuming the distribution of noise is tractable, such as additive white Gaussian noise (AWGN), the near-optimal detected signal can be obtained by linear detectors, which is a commonly used model-based detector, including the minimum mean squared error (MMSE) detector \cite{rusek2012scaling}. However, in actual communications, noise composition is extremely complex, usually including interference between users, interference from non-users, and the influence of signals in the same frequency band. This intricate feature results in the system usually having no prior knowledge of the noise signal $p_n$, thereby increasing signal detection error \cite{daha2024optimizing}. 
Therefore, it is crucial to obtain reliable MLE to estimate the distribution of signals and noise accurately 
in wireless communication systems, which in turn enhances communication quality. 
Additionally, the necessity for distributed learning makes GenAI methods particularly promising options. For example, the maximum Normalizing Flow Estimate method (MANFE) is a GenAI model utilizing normalizing flow to detect multiple-input-multiple-output (MIMO) signals, demonstrating excellent performance in scenes with unknown noise \cite{he2021learning}. 
B. Fesl et al. proposed a diffusion-based MIMO channel estimation method utilizing the generated content as generative priors~\cite{fesl2024diffusion}.

However, using the DM does not guarantee that the generated result is the maximum-likelihood estimation point. 
Inspired by Diffusion Classifier \cite{li2023your}, using the diffusion process instead of denoising process, 
this paper proposes the MLEI framework, which aims to solve the maximum-likelihood estimation problem in wireless communications, especially for signal detection.
The contributions of this paper are summarized as follows:
\begin{itemize}
    \item We present a Maximum-Likelihood Estimation Inference (MLEI) framework in the discrete state space to obtain the maximum-likelihood estimation.
    \item We consider an uplink transmission of a near-filed MIMO communication system, 
    where the channel gain matrix is a function of both the distance and direction.
    Then, we conduct several experiments to show how the MLEI framework works in these complex communication scenarios.
To the best of our knowledge, this is the first attempt to use generative models for signal detection in near-field communication scenarios.
\end{itemize}

\section{Maximum-Likelihood Estimation Inference}


In this section, we propose our Maximum Likelihood Estimation Inference (MLEI) framework in the discrete state space. First, we briefly review the fundamental concepts, formulations, and training framework of the diffusion model.

\subsection{Denoising Diffusion Model}

To accurately estimate the unknown noise distribution, we intend to use the diffusion model (DM), a generative model that excels in many complex distribution generation tasks, to detect the signal \cite{wang2024generative}.
The DMs consist of two main processes: the forward diffusion process and the denoising process \cite{ho2020denoising}. The forward diffusion process adds noise to the data, while the denoising process removes the noise to restore the data.
Specifically, consider a data distribution $\bm{n}_0\sim p(\bm{n})$, the forward diffusion process adds $T$ Gaussian noises hierarchically to make the original distribution $p(\bm{n})$ converge to the standard normal distribution.
Formally, The disturbance data $\bm{x}_t$ with $t$ times Gaussian noise added is expressed as,
\begin{equation}
\begin{split}
    \bm{n}_t &= \sqrt{\gamma_t} \bm{n}_{t-1} + \sqrt{1-\gamma_t} \bm{\epsilon}_{t-1}\\
    & = \sqrt{\overline{\gamma}_t}\bm{n}_0 + \sqrt{1-\overline{\gamma}_t}\overline{\bm{\epsilon}}_t,
\end{split}
\end{equation}
where $\overline{\gamma}_t = \prod_{i=1}^t \gamma_i$ and $\gamma_t\in (0, 1)$ is the hyperparameter for adjusting the mean and variance of the $t$-th noise; $\epsilon_{t-1}$ and $\overline{\epsilon}_t$ are added noise all following the standard normal distribution, i.e., $\bm{\epsilon}_{t-1}$, $\overline{\bm{\epsilon}}_t\sim \mathcal{N}(0,\bm{I})$, where $\bm{I}$ denotes the identity matrix. When setting hyperparameters with $\overline{\gamma}_T\approx 0$ and $1-\overline{\gamma}_T\approx 1$, the perturbed data $\bm{n}_T$ with $T$ times Gaussian noises added will approximately follow the standard normal distribution, i.e., $\bm{n}_T\sim \mathcal{N}(0,\bm{I})$.

As the opposite of the forward diffusion process, the denoising process aims to recover the pure data from the noise distribution. 
Since forward diffusion and denoising are inverse Markov chains in the same representative form, the denoising transitions can be expressed as \cite{sohl2015deep}
\begin{equation}
\label{eq:deno1}
    q(\bm{n}_{t-1}\vert \bm{n}_t) = \mathcal{N}(\bm{n}_{t-1}; \bm{\mu}_t, \bm{\sigma}_t^2 \bm{I}).
\end{equation}
Generally, the denoising transitions in Eq. \eqref{eq:deno1} are challenging to derive and compute directly. Thus, a noise neural network is trained to learn the transition distribution through the variational inference principle \cite{blei2017variational}. By the Bayes’ theorem, the learned transitions can be represented as
\begin{align}
     q_{\theta}(\bm{n}_{t-1}\vert\bm{n}_t, \bm{n}_0) &= \mathcal{N}(\bm{n}_{t-1};\bm{\mu}_{\theta}(\bm{n}_t,t), \bm{\sigma}_{t}^{2}\bm{I}),\label{eq:tri1}\\
     \bm{\mu}_\theta (\bm{n}_t,t)&= \frac{1}{\sqrt{\gamma_t}}(\bm{n}_t - \frac{1-\gamma_t}{\sqrt{1-\overline{\gamma}_t}}\bm{\epsilon}_\theta(\bm{n}_t,t)),\label{eq:inference}\\
     \bm{\sigma}_t^2 &= \frac{(1-\gamma_t)(1-\overline{\gamma}_{t-1})}{1-\overline{\gamma}_t},
\end{align}
where $\bm{\epsilon}_{\theta}(\bm{n}_t,t)$ is the noise neural network.
The DM's training aims to maximize the evidence lower bound (ELBO) on the log-likelihood of the original distribution $\log p(\bm{n})$ \cite{fesl2024diffusion}.
Given a trained noise network, new samples can be generated via sampling from estimated transitions, which is
\begin{equation}
\label{eq:sample}
    \bm{n}_{t-1} = \frac{1}{\sqrt{\gamma_t}}(\bm{n}_t - \frac{1-\gamma_t}{\sqrt{1-\overline{\gamma}_t}}\bm{\epsilon}_\theta(\bm{n}_t,t)) + \bm{\sigma}_t \bm{\epsilon}_{t},
\end{equation}
where $\bm{\epsilon}_t\in\mathcal{N}(0,\bm{I})$, $t=T,T-1,\dots,1$, and $\bm{n}_T\in\mathcal{N}(0,\bm{I})$ is the initial noise. 
The sampling result $\bm{n}_0$ can be considered as the data from the original distribution $p(\bm{n})$, i.e., $\bm{n}_0\sim p(\bm{n})$.

In the training process, given the input $\bm{n}_0\in \mathcal{D}$,  we perturb it via Gaussian noise to obtain the transition distribution $p({\bm{n}}_t\vert{\bm{n}}_{t-1}) = \mathcal{N}({\bm{n}}_t; \sqrt{\gamma_{t-1}}{\bm{n}}_{t-1}, (1-\gamma_t)\bm{I})$. To learn the reverse transition $q_\theta({\bm{n}}_{t-1}\vert{\bm{n}}_{t},{\bm{n}}_{0})$ in Eq. \eqref{eq:tri1}, the DMs aims to maximize the ELBO via minimizing the KL-Divergence between $p_n$ and $q_\theta$.
The loss function is as follows \cite{ho2020denoising}:
\begin{equation}
\label{eq:loss_simple}
\begin{split}
    &l(\theta) := \mathbb{E}_{n_0\in \mathcal{D}}\mathbb{E}_{t\in[0,T], \bm{\epsilon}\in\mathcal{N}(0,\bm{I})} \\
    &~~~~~~~~~~~[
    \vert\vert\bm{\epsilon}-\bm{\epsilon}_\theta(\sqrt{\overline{\gamma}_t}{\bm{n}}_{0} + \sqrt{1-\overline{\gamma}_t}\bm{\epsilon},t)  \vert\vert^2].
\end{split}
\end{equation}

\subsection{Maximum-Likelihood Estimation Inference}

In the inference process, given a discrete solution space $ \mathcal{T}$, where $\bm{n}_0\in \mathcal{T}$, our goal is to find the suitable data, which is the MLE or the closest point to the MLE point of data distribution $p(\bm{n})$. However, it is challenging to derive a formal expression for the MLE of data distribution $p$ due to the difficulty of solving the integral calculation of the neural network $\bm{\epsilon}_\theta$.
One possible solution is to use Monte Carlo sampling \cite{hastings1970monte}, which can approximate the MLE point $\bm{n}_{MLE}$ by multiple samplings. Nonetheless, the DM is well-known for its notoriously long sampling times, which results in the cost of much time for multiple samplings. 

Inspired by Diffusion Classifier \cite{li2023your}, we propose the MLEI framework to estimate the MLE point by comparing the KL-divergence with the perturbed distribution instead of directly obtaining it by denoising.
Specifically, the MLE of distribution $p(\bm{n})$ in the discrete solution space $\mathcal{T}$ can be computed by 
\begin{equation}
    \bm{n}_{MLE} = \arg\max_{\bm{n}_0^i\in\mathcal{T}}p(\bm{n}_0^i).
\end{equation}
Similar to the training process of DM, we use the ELBO of $\log q_\theta(\bm{n})$
in place of $\log p(\bm{n})$, i.e.,
\begin{equation}
\begin{split}
    p(\bm{n}_0^i) \propto 
    \exp\{-\mathbb{E}_{t\in[0,T], \bm{\epsilon}\in\mathcal{N}(0,\bm{I})} [\vert\vert\bm{\epsilon}-\bm{\epsilon}_\theta(\bm{n}^i_t,t)  \vert\vert^2]\},
\end{split}
\end{equation}
where $\bm{\epsilon}_\theta({\bm{n}}^i_t,t) $ is the trained DM model, and $\bm{n}^i_0 \in \mathcal{T}$ is $i$-th potential solution. Based on this, we define an error function $E(\bm{n}_i)$ with the set of sampling pairs $\mathcal{S} = \{({t_{s}},{\bm{\epsilon}_s})\}_{s=1}^{S}$, which can be expressed as follows:
\begin{equation}
\label{eq:result}
E(\bm{n}_i) = \frac{1}{S}\sum_{s=1}^{S}\vert\vert{\bm{\epsilon}_s}-
\bm{\epsilon}_\theta(\sqrt{\overline{\gamma}_{t_{i}}}{\bm{n}}^i_{0} + \sqrt{1-\overline{\gamma}_{t_{s}}}\bm{\epsilon}_s,{t_{s}}
)\vert\vert^2. 
\end{equation}
For each possible signal vector $\bm{n}^i_0$ in $\mathcal{T}$, the MLE or the closest point to the MLE point is the vector with the least error $E$ in Eq. \eqref{eq:result}.
The overall MLEI algorithm is shown as Algorithm \ref{alg:1}.

\begin{algorithm}[htp]
\caption{MLEI}
\label{alg:1}
\begin{algorithmic}[1]\footnotesize
    \Require Timestep $T$, coefficient $\{\gamma_t\}_{t=1}^T$, training set $\mathcal{D}$, discrete solution space $\mathcal{T}$, sampling pairs set $\{({t_{s}},{\bm{\epsilon}_s})\}_{s=1}^S$
    \Repeat
        \State Choose $\bm{n}_0 \in \mathcal{D}$, $t\sim$Uniform $(\{1,\ldots,T\})$, and $\bm{\epsilon}\sim\mathcal{N}(0,\bm{I})$
        \State Take gradient descent step on Eq. \eqref{eq:loss_simple}
    \Until converged
    \ForAll{$\bm{n}^i_0\in\mathcal{T}$}:
        \State Compute $E(\bm{n}^i_0)$ in Eq. \eqref{eq:result} 
    \EndFor
    \State ${\bm{n}}_d \leftarrow \arg\min_{\bm{n}^i_0\in\mathcal{T}} E(\bm{n}^i_0)$
    \State \Return ${\bm{n}}_d$
\end{algorithmic}
\end{algorithm}

\section{Case Study}





\subsection{Signal Detection in Near-field MIMO System}

\begin{figure}[htbp]
    \centering
    \includegraphics[width= 0.95\linewidth]{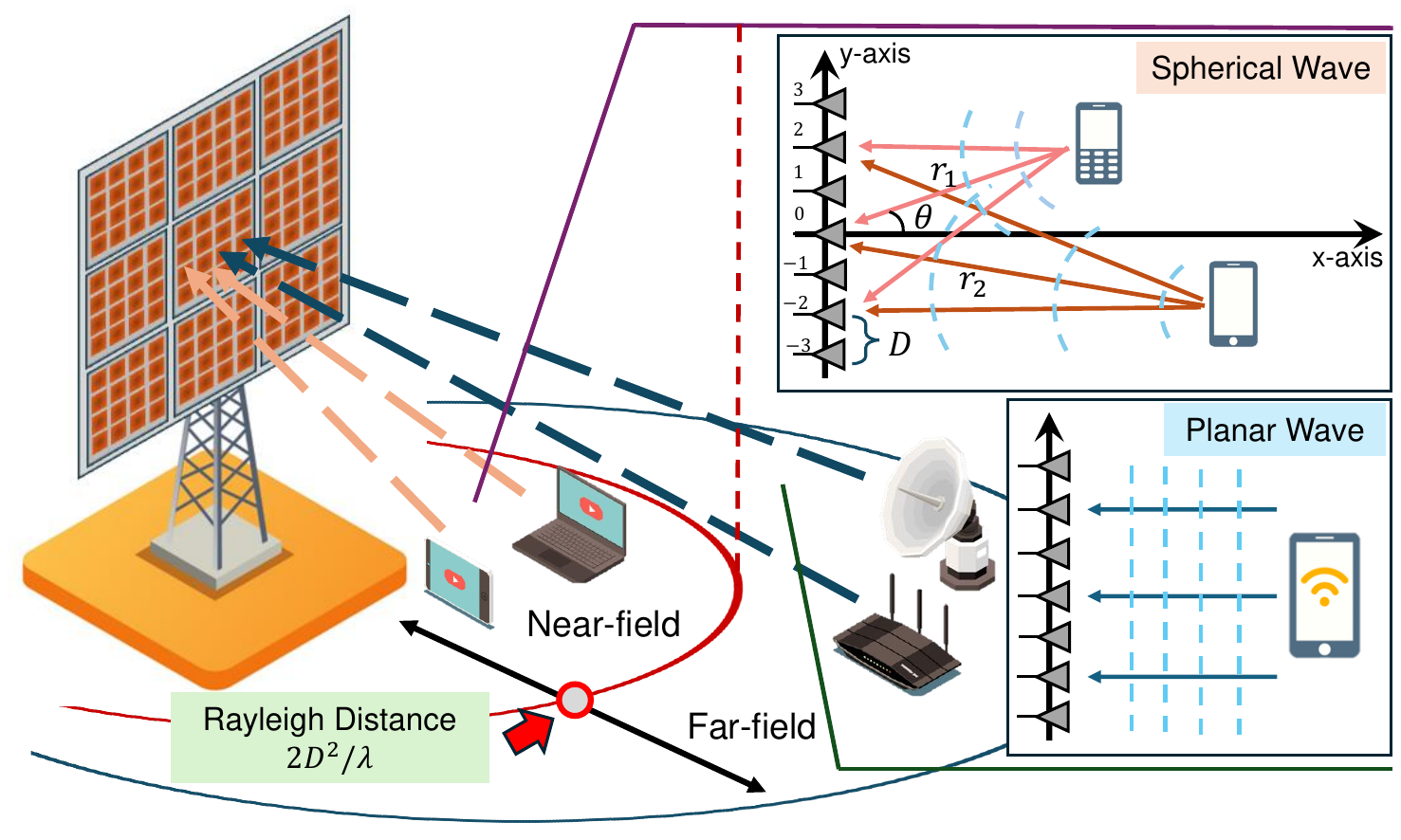}
    \caption{Illustration of the near-field region and the far-field region divided by the Rayleigh distance, with spherical wave and planar wave, respectively.}
    \label{fig:model}
\end{figure}

As shown in Fig. \ref{fig:model}, we consider an uplink transmission of a near-filed MIMO communication system consisting of $M$ mobile terminals (MTs) and a base station (BS) equipped with $N$-antenna uniform linear array. 
Denoting $r_m$ as the distance between the $m$-th MT and the center of the receive array antenna, the MT is located in the Fresnel zone (radiative near-filed) of the array, i.e., $0.62(D^3/\lambda)^{1/2}<r_m<2D^2/\lambda$, where $m = 1,2,\ldots, M$, $D$ denotes array aperture and $\lambda$ is the carrier’s wavelength \cite{zhang2024physical}.
Assuming each MT transmits the source signal to the BS through independent paths, the received signal $\bm{y}$ can be written as follows:
\begin{equation}
\label{eq:o_signal}
    \bm{y} = \bm{A}\bm{x} + \bm{n},
\end{equation}
where $\bm{y} = [y_1, \ldots, y_{N}]^{T}\in \mathbb{C}^{N\times 1}$, $\bm{x} = [x_1,\ldots,x_M]^{T}\in \mathbb{C}^{M\times 1}$ represents the source signal vector, 
$\bm{A}\in\mathbb{C}^{N\times M}$ denotes the channel matrix, 
and $\bm{n}\in\mathbb{C}^{N\times 1}$ is an additive measurement noise composed of the interference by non-user signals, which is independent and identically distributed (i.i.d) with an unknown distribution $p_n(\cdot)$.
The received signal of $n$-th antenna can be expressed as:
\begin{equation}
    y_n = \sum_{i=1}^{M}a_{ni}\cdot x_i + n_i,
\end{equation}
where $a_{ni}$ denotes the element in matrix $A$. 
Here, $M$-user joint detection via MLEI is realized in the presence of the measurement noise and inherent channel estimation errors between $N$ antenna elements and $M$ users.
We consider line-of-sight propagation in the small-coverage system and do not consider the multipath effect.
In near-field communication, the channel matrix $A$ is composed of a series of array steering vectors, where each steering vector is derived from the accurate spherical wave. Specifically, the array steering vector of the $m$-the MT can be expressed as follows \cite{cao2020complex}:
\begin{equation}
\begin{split}
    \bm{S}_{m}(\theta_m,r_m)=& [\kappa_{1,m}\exp\{-j k_c(r_{1,m}-r_m)\},\ldots,\\
    &~~~~~~~~\kappa_{N,m}\exp\{-j k_c(r_{N,m}-r_m)\}],
    \end{split}
\end{equation}
where $\theta_m\in(-1,1)$ denotes the corresponding spatial angle from the $m$-th MT source to the receiver; $k_c=\frac{2\pi}{\lambda}$ represents the wave number; $\kappa_{n,m}= \frac{r_m}{r_{n,m}}$ is the corresponding amplitude; $r_{n,m}$ denotes the distance from the $n$-th array element to the $m$-the MT. 
According to the geometric relationship, we have
\begin{equation}
    r_{n,m} = \sqrt{r_m^2 + \delta_n^2d^2-2r_m\delta_n\theta_md},
\end{equation}
where $\delta_n = \frac{2n-N+1}{2}, n=0,1,\ldots,N-1$, 
$d$ is the element spacing of the antenna array,
and $(0,\delta_nd)$ denotes the coordinate of the $n$-the antenna.

To simulate unknown interference noisy data, we use a model called symmetric $\alpha$-stable (S$\alpha$S) noise \cite{he2021learning}.
Specifically, the probability density function of S$\alpha$S noise can be written as,
\begin{equation}
    f(n) = \frac{1}{2\pi}\int_{-\infty}^{+\infty}e^{-\vert \theta \vert \sigma^\alpha -j\theta n}d\theta,
\end{equation}
where $\sigma>0$ denotes the scale exponent and $\alpha \in (0,2]$ represents the characteristic exponent. 
When $\alpha\in(1,2)$, the S$\alpha$S noise does not have a closed-form expression. This means that we can only estimate $f(n)$ using numerical simulations and cannot compute the candidate probabilities from Eq. \eqref{eq:mle}, which shows the unknown noise characteristic we anticipated. Additionally, when $\alpha=1$, the S$\alpha$S noise follows a Cauchy distribution and a Gaussian distribution when $\alpha=2$.

\subsection{Network Architecture}

In the case study, wireless signals are converted to the
real-valued representation to express the complex-valued input equivalently, i.e.,
\begin{equation}
\overbrace{
\begin{bmatrix}
Re(\bm{y}) \\
Im(\bm{y}) \\
\end{bmatrix}}^{\overline{\bm{y}}} = 
\overbrace{
\begin{bmatrix}
Re(\bm{A}) &-Im(\bm{A}) \\   
Im(\bm{A}) & Re(\bm{A}) \\
\end{bmatrix}}^{\bm{\overline{A}}}
\overbrace{
\begin{bmatrix}
Re(\bm{x}) \\
Im(\bm{x}) \\
\end{bmatrix}}^{\overline{\bm{x}}} + 
\overbrace{
\begin{bmatrix}
Re(\bm{n}) \\
Im(\bm{n}) \\
\end{bmatrix}}^{\overline{\bm{n}}},
\end{equation}
where $Re(\cdot)$ and $Im(\cdot)$ represent the real and imaginary part functions, respectively. 
Based on this, the received signal (Eq. \eqref{eq:o_signal}) can be reformulated as follow:
\begin{equation}
\label{eq:signal}
    \overline{\bm{y}} = \bm{\overline{A}\overline{x}} + \bm{\overline{n}}.
\end{equation}



In the deployment, we first collect 
the transmitted signal ${\bm{x}}$, received signal ${\bm{y}}$, and channel information ${\bm{A}}$ in the given unknown noise scenario to establish a signal dataset $\mathcal{D} = \{({\bm{x}}^i, {\bm{y}}^i, {\bm{A}}^i)\}_{i=1}^{\vert\mathcal{D}\vert} = \{\bm{n}^i_0 | \bm{n}^i_0 = \bm{y}^i-\bm{A}^i\bm{x}^i\}_{i=1}^{\vert\mathcal{D}\vert}$.
Then, the DM network $\bm{\epsilon}_{\theta}({\bm{n}}_t, t)$ is trained to estimate the unknown signal noise distribution $p_n$.

\begin{figure*}[htbp]
    \centering
    \includegraphics[width= 0.7\linewidth]{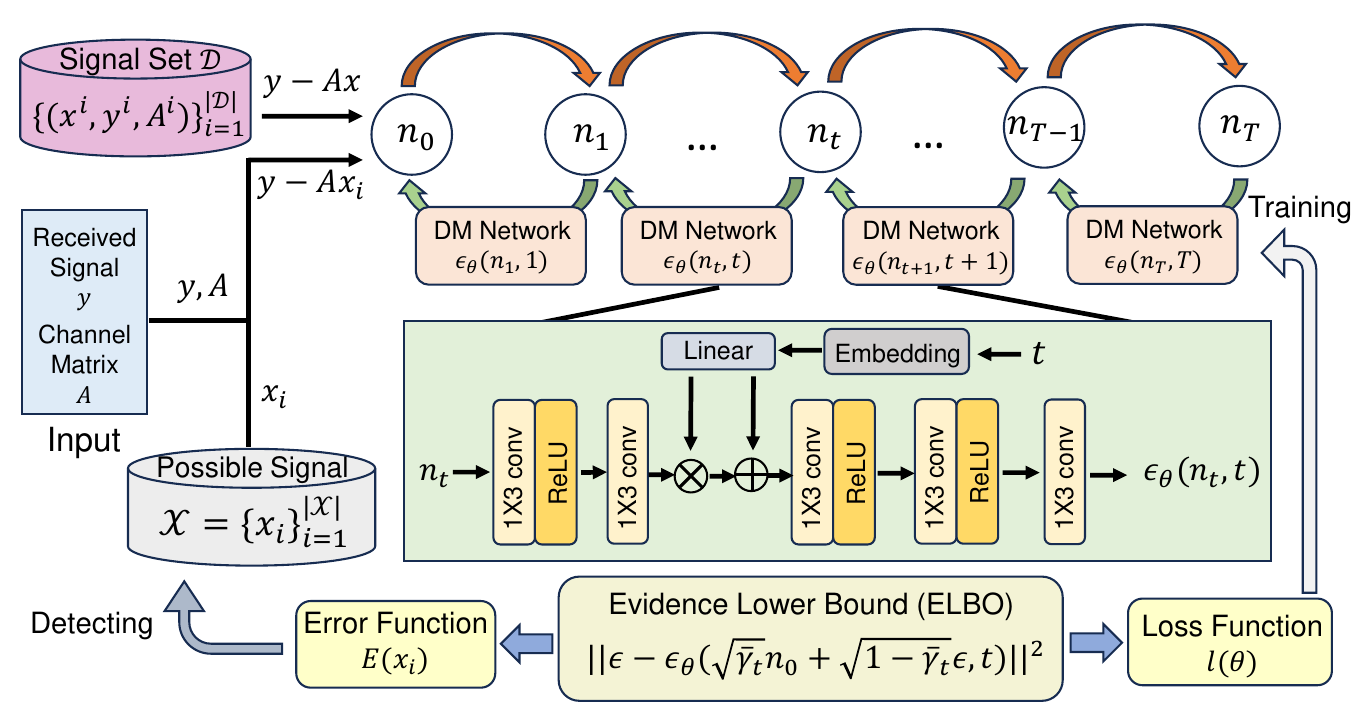}
    \caption{The overall architecture of MLEI.}
    \label{fig:framework}
\end{figure*}

As depicted in Fig. \ref{fig:framework}, the DM network utilizes two 1D convolution layers with a kernel size $k$ of 3 and rectified linear unit (ReLU) activation to progressively increase the number of channels to $C_{max}$.
The embedded time $t$ is split into a scaling part and a bias part and then is connected to the output of convolution layers.
Afterward, three 1D convolution layers are utilized by gradually decreasing the convolutional channels into the initial channel $C_{init}$.
The same linear schedule of $\gamma_t$ is chosen as in \cite{fesl2024diffusion}.

\section{Simulation Results}

In this section, we conduct simulations to showcase the effectiveness of the proposed MLEI detection framework.
All experiments were conducted on an A100 Linux server using Python and the PyTorch package. The main program is based on the DM framework in \cite{fesl2024diffusion}.








\subsection{Environment Set}

We consider the number of MTs and antennas of the BS as $(M,N) = (4,5)$ in the near-field MIMO scenario.
The signals transmitted have a carrier frequency of $f = 28$ GHz, corresponding to a wavelength of $\lambda = 0.0107$ m. We set the array spacing as $d = \lambda/2$. In this case, the Fresnel region is between $112\lambda$ and $2048\lambda$ \cite{cao2020complex}.
The modulation scheme is quadrature phase shift keying (QPSK) with $P= 4$. For the training set, we assume the receiver has perfect knowledge of the channel state information (CSI).
We set $\sigma =1$ and choose the value of $\alpha$ to be within the $[1, 2]$ range for the signal-to-noise ratio (SNR) from 0 dB to 25 dB to test the performance of different detectors.
We leverage BERs as performance metrics to evaluate the accuracy of signal detection.
In the training process, 
we generated the training dataset with $|\mathcal{D}| = 90000$, and the size of the corresponding testing dataset is $10000$. 
For MLEI, we choose a denoising time-step $T=10$ with the same diffusion noise schedule in \cite{fesl2024diffusion} and set the initial number of channels $C_{init}=1$ and the maximum value $C_{max}=64$ for the DM channels.
During the detection process, 
we choose a sampling pairs shuddering as $\mathcal{S} = \{(s,\bm{0})\}_{s=1}^{10}$ to ensure the stability of detection.

\subsection{Baseline Approach}

We compare the proposed MLEI with the
following classical and learning-based baselines.
The classical Euclidean distance-based MLE detector following $\bm{x}_{MLE}=\arg\min_{\bm{x}\in\mathcal{X}}\vert\vert\bm{y}-\bm{A}\bm{x}\vert\vert^2$ is denoted as ``MLE", which is equivalent to minimum distance decision rule when noise is Gaussian. Additionally, two model-based linear detectors based on MMSE with variance as 1 and Zero-Forcing (ZF) are denoted as ``MMSE" and ``ZF", respectively  \cite{rusek2012scaling}.
A learning-based signal detection framework DetNet from \cite{samuel2019learning} with 30 layers is represented as ``DetNet". 
Additionally, we compare the proposed MLEI with another generative model-based method labeled ``MANFE" with the same setting in \cite{he2021learning}.
Specifically, each flow step consists of an activation normalization layer, an invertible $1\times 1$ convolution, and an alternating affine coupling layer.

Furthermore, we evaluate the impact of different denoising steps with $T=5,30$ and compare the proposed MLEI to the original diffusion model with $T=10$ and 20 times sampling.
We finally compare the performance on the testing set of imperfect CSI (ICSI) with the same SNR estimation error in the estimated channel matrix.
All training approaches are trained for 500 epochs with an early stopping mechanism.

\begin{figure*}[htbp]
\centering
\begin{subfigure}{.48\textwidth}
  \centering
  \includegraphics[width=0.65\linewidth]{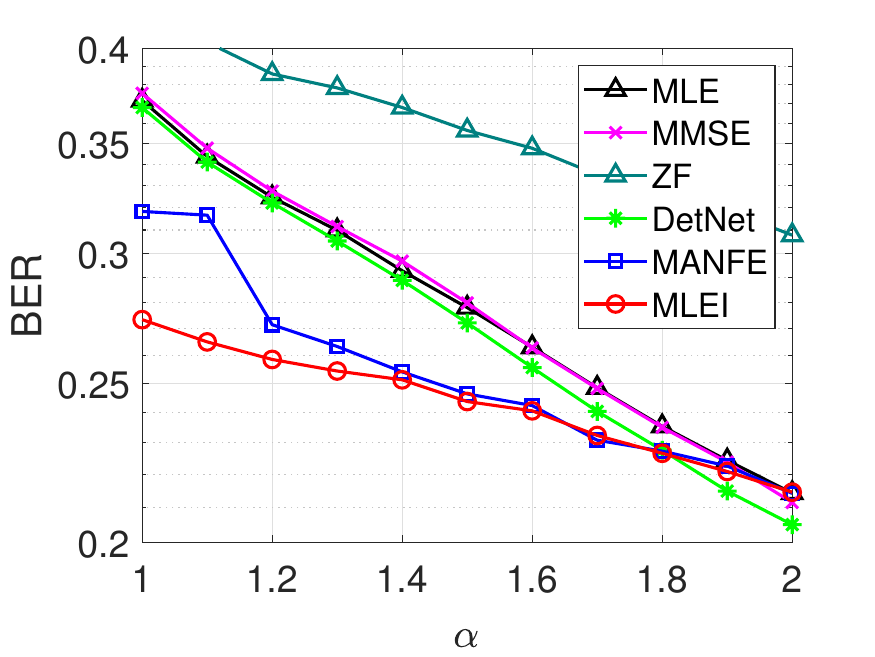} 
  \caption{BER vs $\alpha$ with SNR = 5 dB}
  \label{fig:sub1}
\end{subfigure}%
\begin{subfigure}{.48\textwidth}
  \centering
  \includegraphics[width=0.65\linewidth]{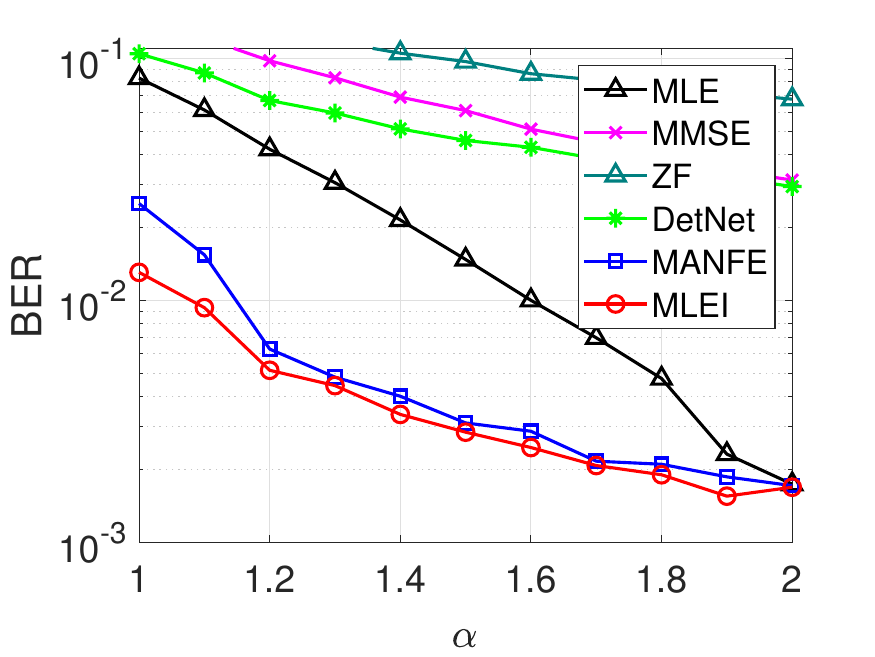}
  \caption{BER vs $\alpha$ with SNR = 25 dB}
  \label{fig:sub2}
\end{subfigure}
\\
\begin{subfigure}{.48\textwidth}
  \centering
  \includegraphics[width=0.65\linewidth]{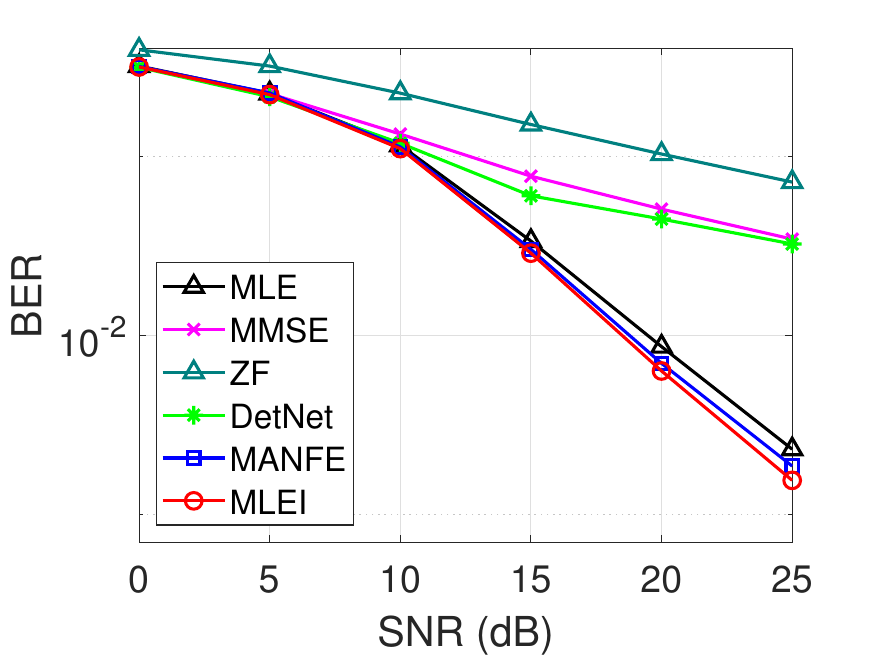}
  \caption{BER vs SNR with $\alpha$ = 1.9}
  \label{fig:sub4}
\end{subfigure}
\begin{subfigure}{.48\textwidth}
  \centering
  \includegraphics[width=0.65\linewidth]{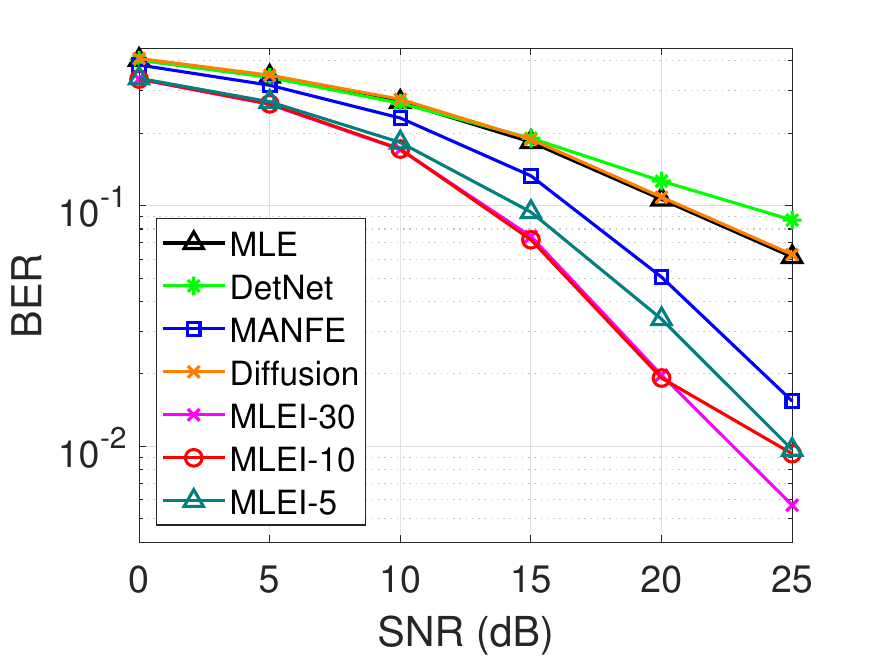}
  \caption{BER vs SNR with $\alpha$ = 1.1}
  \label{fig:sub3}
\end{subfigure}%
\caption{BER performance comparison curves for $4$ methods in
$4\times5$ near-field MIMO system.}
\label{fig:test}
\end{figure*}

\subsection{Performance Evaluation}

Fig. \ref{fig:test} assesses the BER performance for the proposed MLEI method and other baselines in various experimental settings.
First, we examine the BER performance of various methods under SNRs 5 dB and 25 dB.
Fig. \ref{fig:sub1} illustrates that in low SNR scenarios, the proposed MLEI method outperforms or has competitive performance with other methods.
Specifically, when $\alpha=1.1$, where extreme values are more likely to occur, the MLEI can significantly reduce the BER of MLE, DetNet, and MANFE to about 22.9\%, 22.3\%, and 16.3\% respectively.
Additionally, when the noise approaches a Gaussian distribution with $\alpha=2$, the performance improvement becomes less significant and the performance of MLEI is even worse than DetNet. When $\alpha>1.8$, although the performance is not as good as that of DetNet, it still exhibits slightly lower error rates than the existing generative learning method MANFE.
On the other hand, when SNR is 25 dB in Fig. \ref{fig:sub2}, MLEI achieves the best performance on all impulsive noise levels. Even compared with MANFE, which has the closest performance, 
MLEI reduces the average BER to about 31.1\%.

\begin{figure}[htbp]
\centering
  \includegraphics[width=0.65\linewidth]{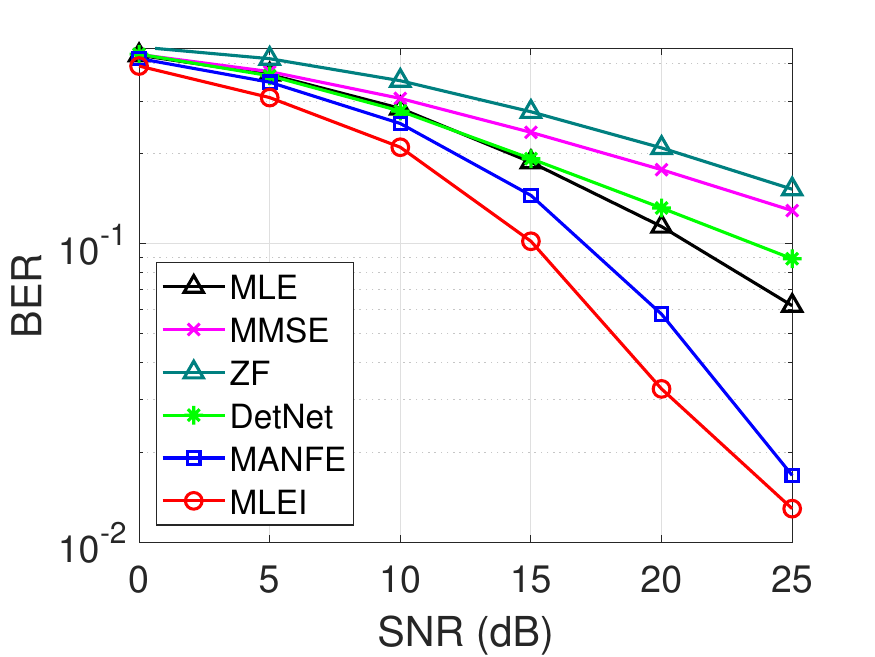}
  \label{fig:icsi}
\caption{BER performance comparison curves with ICSI.}
\label{fig:test2}
\end{figure}

Then, we investigate the BER performance under complex noise distribution where $\alpha$ is $1.9$ and $1.1$, respectively, as shown in Figs. \ref{fig:sub4} and \ref{fig:sub3}.
In Fig. \ref{fig:sub4}, it is demonstrated that even for noise close to a Gaussian distribution, MLEI performs similarly to other methods, and MLEI shows significant improvement as the SNR increases. Specifically, at 25 dB SNR, MLEI outperforms MLE by 32.97\% and MANFE by 16.78\%.
When the denoising step size is set to 5, 10, and 30, corresponding to ``MLEI-5", ``MLEI-10", and ``MLEI-30", respectively, all MLEI models display a stronger learning ability for complex unknown noise distributions across all SNR conditions. As the denoising step size increases, the BER performance of MLEI improves. Particularly at an SNR of 25 dB, MLEI-30 achieves a BER of 0.57\%, reducing the BER of the baseline generative model MANFE by approximately 63.05\%.
Additionally, 
compared with the numerical method, MLE, and learning method, DetNet, MLEI, the BER of MLEI is significantly reduced, and as the SNR increases, the reduction is more obvious. 
Furthermore, by directly leveraging generated signals of the diffusion model, the BER performance, represented as ``Diffusion", is almost consistent with the performance of MLE. This performance gap reflects the effectiveness of the new error function proposed in the inference process. Finally, the BER performance comparison under $\alpha=1.1$ and ICSI is presented in Fig. \ref{fig:test2}. Under ICSI conditions, MLEI demonstrates the best performance. Specifically, the BER performance of MLEI decreases the BER of MANFE and MLE by an average of 14.31\% and 26.60\%, respectively. This performance gap shows that our algorithm is more robust under disturbance conditions.



Additionally, the computational complexities of the online detection algorithms are shown in Table \ref{tab:complexity}. The complexity of the MLEI algorithm is \(\mathcal{O}((|\mathcal{S}|C_{\text{max}}kM + MN)P^{N})\). For the MLE and MANFE, the complexity are given by \(\mathcal{O}(MNP^{N})\), \(\mathcal{O}((KM + MN)P^N)\), respectively, where \(K\) represents the depth of the MANFE algorithm, and \(k\) denotes the kernel size used in the MLEI method. Additionally,
As the number of MTs $M$ and the number of antennas $N$ increase, the complexity of the three MLE-based methods, including MLE, MANFE, and MLEI, is of the same magnitude. 
Although model-based detectors, including MMSE, ZF, and DetNet, have lower complexity at the polynomial level, they require assumptions about the noise distribution, leading to worse performance in the case of unknown noise.



In summary, the proposed MLEI framework has a strong ability to handle unknown distribution noise 
and has better performance than existing methods with the same magnitude of complexity as the noise becomes more complex. 
Moreover, in all near-field communication qualities, especially in scenarios with good communication quality, it still has comparable performance to those of the other methods.


\begin{table}[h]\small
\centering
\caption{Computational Complexity of Detection Algorithms}
\label{tab:complexity}

\begin{tabular}{c|c}
\toprule
\textbf{Abbreviations} & \textbf{Complexity} \\ \hline
MLE & $\mathcal{O}(MNP^{N})$ \\ \hline
MANFE & $\mathcal{O}((KM+MN)P^N)$\\ \hline
MLEI & $\mathcal{O}((|\mathcal{S}|C_{max}kM+MN)P^{N})^1$ \\ \hline
DetNet & $\mathcal{O}(I(N^2+ L_h(3N + L_z))^2$ \\ \hline
MMSE & $\mathcal{O}(N^3)$ \\ \hline
ZF & $\mathcal{O}(N^3)$ \\ 
\bottomrule
\end{tabular}
\scriptsize 
\begin{flushleft}
$^1$ $K$ represents the depth of the flow network and $k$ is the kernel size of MLEI.\\
$^2$ $L_z$ and $L_h$ are the size of a latent parameter and the size of hidden layers, respectively; $I$ denotes the number of iterations \cite{samuel2019learning}.
\end{flushleft}
\end{table}
\vspace{-0.1cm}
\section{Conclusion}

In this paper, we have presented a Maximum-Likelihood Estimation Inference (MLEI) framework in the DM to solve the maximum-likelihood estimation point of unknown data distribution in discrete space. By employing the proposed framework in a 
near-field signal detection scenarios, 
the simulation results have demonstrated that the proposed MLEI framework outperforms the classical numerical method and existing learning-based methods, especially when the unknown distribution is more intricate.


\bibliography{Ref}

\end{document}